\documentclass[prc,aps,preprint]{revtex4-1}
\usepackage{CJK}
\usepackage{mathrsfs}
\usepackage{amssymb}
\usepackage{graphicx}
\usepackage{bm}
\usepackage{graphicx}
\usepackage{float} 
\usepackage{longtable}
\usepackage{amsmath}
\usepackage{color}
\usepackage{multirow}
\usepackage{rotating}
\usepackage{upgreek}
\newfont{\largemi}{cmmi10}
\newfont{\smallmi}{cmmi6}

\def\eqref#1{Eq.~(\ref{eq:#1})}

\makeatletter

\begin{document}
\setlength{\abovedisplayskip}{1ex} 
\setlength{\belowdisplayskip}{1ex} 

\title{Transition sum rules in the shell model}

\author{Yi Lu }
\affiliation{College of Physics and Engineering, Qufu Normal University, 57 Jingxuan West Road, Qufu, Shandong 273165, China}
\affiliation{School of Physics and Astronomy, Shanghai Jiao Tong University, Shanghai 200240, China}
\affiliation{Department of Physics, San Diego State University, 5500 Campanile Drive, San Diego, CA 02182-1233, United States}
\author{Calvin W. Johnson}
\affiliation{Department of Physics, San Diego State University, 5500 Campanile Drive, San Diego, CA 02182-1233, United States}
\date{\today}
\begin{abstract}
An important characterization of electromagnetic and weak transitions in atomic nuclei
 are sum rules. 
We focus on the non-energy-weighted sum rule (NEWSR), or total strength, and the energy-weighted sum rule (EWSR);
the ratio of the EWSR to the NEWSR is the
 centroid or average energy of transition strengths from an nuclear initial state to all  allowed final states.
 These sum rules can be expressed as expectation values of operators, in the case of the EWSR a  double commutator. 
 While most prior applications of the double-commutator have been 
to special cases,
we  derive general formulas for matrix elements of both operators in a shell model framework (occupation space),
given the input matrix elements for the nuclear Hamiltonian and for the transition operator.
With these new formulas, we easily evaluate centroids of transition strength functions, with no need to calculate daughter states.
We apply this simple tool to a number of nuclides, and demonstrate the sum rules follow smooth secular behavior as a function 
of initial energy, as well as compare the electric dipole (E1) sum rule against the famous
Thomas-Reiche-Kuhn version. 
 We also find surprising systematic behaviors for  ground state electric quadrupole (E2) centroids in the $sd$-shell.
\end{abstract}
\maketitle

\section{Introduction}
\label{sec1}
Atomic nuclei are neither static nor exist in isolation. Their transitions play important roles 
in fundamental, applied, and astro-physics, as well as revealing key information about nuclear structure beyond just excitation energies. In this paper we focus on electromagnetic and weak transitions; such
transition strength distributions are important for $\gamma$-spectroscopy, nucleosynthesis and $\beta\beta$ decays, as they are used to extract level densities \cite{PhysRevC.79.024316}, calculate nuclear reaction rates in stellar processes \cite{PhysRevC.82.014318} and analyze $\beta\beta$ decay matrix elements \cite{PhysRevLett.103.012503}.

The strength function for a transition operator $\hat{F}$ from an initial state $i$ at energy $E_i$, to a final state $f$ at absolute energy $E_f$ and 
excitation energy $E_x = E_f - E_i$ is defined as
\begin{equation}
\label{strength}
S(E_i, E_x) = \sum_f \delta(E_x + E_i - E_f ) \left | \langle f  \left | \hat{F}  \right | i \rangle \right |^2.
\end{equation}
Sum rules are moments of the strength function,
\begin{equation}
\label{sum_rules}
S_k(E_i) = \int \left ( E_x \right )^k S(E_i, E_x) \, dE_x.
\end{equation}
 Two of the most important sum rules, which we consider here, are  $S_0$, the non-energy-weighted 
sum rule (NEWSR) or total strength,  and $S_1$, the energy-weighted 
sum rule (EWSR).   These sum rules provide compact information about strength functions. For example, the 
famous Ikeda sum rule \cite{ikeda1963p} for Gamow-Teller (GT) transitions is the difference between the total $\beta-$ strength and total $\beta+$ strength:
$$
S_0(GT_-) - S_0(GT_+)= 3(N-Z)g_A^2,
$$
where $g_A$ is the axial vector coupling relative to the vector coupling $g_V$.
For investigations of `quenching' of $g_A$ \cite{RevModPhys.64.491}, the NEWSR 
 $S_0$ can be a probe of the missing strengths due to hypothesized cross-shell configurations.

The centroid of a strength distribution is just the ratio of the EWSR to the NEWSR,
\begin{eqnarray}
E_{\rm centroid} (E_i) = \frac{ S_1(E_i)}{S_0(E_i)}.
\label{E_centroid}
\end{eqnarray}
For a compact distribution of a giant resonance, $E_{\rm centroid}(E_i)$ will be roughly the location of the resonance peak, relative to the parent state energy $E_i$; of course, in the case of highly fragmented strength distributions this interpretation no longer holds, and in severely
truncated model spaces the centroid will be too low compared to experiment. 
Both the NEWSR $S_0$ and $E_{\rm centroid} (E_i)$ can test  the validity of the general Brink-Axel hypothesis \cite{PhysRevLett.116.012502, Johnson2015Systematics}.
The general Brink-Axel hypothesis \cite{Brink1955thesis, PhysRev.126.671, Brink2009talk} assumes that the strength distribution of transitions from any parent state is approximately the same, thus as a result $E_{\rm centroid}(E_i)$ is independent on $E_i$.
Though it seems this hypothesis needs to be modified for E1\cite{PhysRevLett.70.533,PhysRevLett.74.3748,PhysRevC.86.051302}, M1\cite{PhysRevC.76.044303,PhysRevLett.111.232504,PhysRevLett.113.252502} (the low-energy $\gamma$ anomaly) and GT\cite{PhysRevC.90.065808} transitions, it is still being widely used to calculate neutron-capture rates \cite{PhysRevC.91.044318}, extract nuclear level densities \cite{PhysRevC.68.054326, PhysRevC.79.024316, PhysRevC.74.014314} and can have a substantial impact on astrophysical relevance \cite{Goriely1998Radiative, PhysRevC.82.014318}.

Sum rules are appealing not only because they characterize strength functions, but also because using closure some sum rules can be rewritten as expectation values of 
operators \cite{ring2004nuclear}.  
Allowing for transition operators with good angular momentum rank $K$, one should sum over the $z$-component $M$,
and the total strength 
$S_0(E_i)$ becomes
\begin{equation}
\sum_f \sum_M |\langle f | \hat{F}_{K, M} | i \rangle|^2 =\sum_M \langle i | (\hat{F}_{K,M})^\dagger \hat{F}_{K,M} | i \rangle.
\label{S_0}
\end{equation}
Thus $S_0(E_i)$ can be easily evaluated numerically without calculating any final state. 
The strength sum can be used to evaluate the former mentioned Ikeda sum rule,  useful as a check  on computations.

The EWSR can  be written as the expectation value of a double commutator, as long as the transition operator behaves as a spherical 
operator under Hermitian conjugation \cite{Edmonds1996Angular},
\begin{equation}
\left( \hat{F}_{KM} \right)^\dagger = (-1)^{M} \hat{F}_{K,-M}.
\label{sphtensoradjoint}
\end{equation}
If we do not have (\ref{sphtensoradjoint}), one cannot write the EWSR operator as a double commutator.
The requirement of this will have consequences when we look at charge-changing transition such as $\beta$ decay. In that case, one must 
include both $\beta-$ and $\beta+$ transitions.

Invoking closure and Eq.~(\ref{sphtensoradjoint}), $S_1(E_i)$ becomes 
\begin{equation}
 \Big\langle i \Big| \frac{1}{2}\sum_M (-1)^M \left[ \hat{F}_{K,-M}, [\hat{H}, \hat{F}_{K,M}] \right] \Big| i \Big\rangle.
\label{S_1}
\end{equation}
As an example, the Thomas-Reiche-Kuhn sum rule \cite{towner1977shell} evaluates the energy-weighted sum of $E1$ strengths of an atom with $N$ electrons, and conserves to a constant proportional to $N/m_e$.
In nuclear physics the corresponding sum rule is similar, though the EWSR is proportional to $NZ/2 Am_N$ because the dipole is relative to the center of mass.
Another example is related with the ``scissor mode" in rare-earth nuclei \cite{RevModPhys.82.2365}, for which the EWSR of low-lying ( $< 4$ MeV ) orbital M1 transitions shows a striking correlation with the $E2$ transition,
\begin{eqnarray}
\sum\limits_x B(M1; 0^+_1 \rightarrow 1^+_x) E_{1^+_x} \propto \sum\limits_x B(E2; 0^+_1 \rightarrow 2^+_x).
\label{M1_EWSR}
\end{eqnarray} 
This EWSR is derived both in the IBM-2 model \cite{PhysRevC.44.R2262}, and in the shell model \cite{PhysRevC.46.2106,PhysRevC.47.2604} with phenomenological interactions.

One can  compute sum rules with the Lanczos algorithm 
\cite{whiteheadmomentmethods,bloom1984gamow,RevModPhys.77.427,PhysRevC.89.064317}, which has a deep connection to the classical moment problem.  Given some initial state $|\Psi_i \rangle$, one applies an transition operator $\hat{F}$ and then uses $\hat{F}|\Psi_i\rangle$ 
as the pivot or starting state in the Lanczos algorithm.
This requires, however, one being able to carry out a matrix-vector multiplication in the Hilbert space under consideration, 
which may not aways be possible or practical, for example in the case of coupled clusters \cite{PhysRevLett.111.122502} or generator coordinate calculations \cite{ring2004nuclear,PhysRevC.90.031301,PhysRevC.96.054310}.
Furthermore, for example in the $M$-scheme, or fixed $J_z$, basis for the configuration-interaction shell model, if the initial state has 
angular momentum $J _i>0$, then applying an operator $\hat{F}_K$ with angular momentum rank $K$ will produce a state with 
mixed $J_f$, with $|J_i-K| \leq J_f \leq J_i+K$ by the triangle rule.  To compare to experiment, however, one generally needs 
a sum over final $M$ values and average over initial $M$ values, and to correctly use the Lanczos method one must either 
do this explicitly or project out states of good angular momentum and extract strength functions via  appropriate Clebsch-Gordan coefficients.
 This point is not emphasized in the literature.

In this paper we go beyond specific cases and, in the next section, write down the general form of the operators (\ref{S_0}) and (\ref{S_1}) in a spherical shell model basis. 
Although straightforward, the EWSR in particular is somewhat involved and to the best of our knowledge not published. 
Appendix A provide some of the details of derivation.  
In Ref. \cite{PandasCommute} we make available a C++ code to generate those operator matrix elements.
With such machinery one can directly compute the 
NEWSR and EWSR easily for many nuclides and many transitions.  
Prior work showed that the NEWSR follows simple secular behavior with the 
initial energy $E_i$ and gave a general argument \cite{Johnson2015Systematics}.  In section \ref{results} we show a few cases and also find simple secular 
behavior.  Finally, we illustrate the applicability by looking at systematics of ground state E1 and E2 sum rules. 

\section{Formalism and formulas}
\label{sec2}

We work in the configuration-interaction shell model, using the occupation representation \cite{Fetter2003Quantum} with fermion single-particle creation and annihilation operators 
$\hat{a}^\dagger$, $\hat{a}$, respectively.  As is standard, our operators have good angular momentum. The labels of each single-particle state 
include the magnitude of angular momentum $j$ and $z$-component $m$; there are other important quantum numbers, in particular parity, orbital 
angular momentum $l$ and label $n$ for the 
radial wave function, but those values are absorbed into the values of matrix elements, so, for example, the details of our derivation are independent 
of whether or not one uses harmonic oscillator or Woods-Saxon or other single-particle radial wave functions.  Because we are working in a shell model basis, we differentiate between 
single-particle \textit{states} (labeled by $j$,$m$, and $l, n, \ldots$) and \textit{orbits}, by which we mean the set of $2j+1$ states with the same $j$ but 
different $m$. We assign fermion operators of different orbits different lower-case Latin letters: $\hat{a}^\dagger$, $\hat{b}^\dagger$, etc., to prevent a 
proliferation of subscripts.  (In our derviations, when discussing generic operators, which may be single-fermion operators or composed of products and sums 
of operators, we use lower-case Greek letters: $\alpha, \beta, \ldots.$)  In order to make our results broadly usable, we will be slightly pedantic.

To denote generic operators $\hat{\alpha}, \hat{\beta}$ coupled up to good total angular momentum $J$ and total $z$-component $M$, we use 
the notation $( \hat{\alpha} \otimes \hat{\beta} )_{JM}$. 
Hence we have the general pair creation operator
\begin{eqnarray}
\hat{A}^\dagger_{JM} (ab) = (\hat{a}^\dagger \otimes\hat{b}^\dagger)_{JM},
\label{Adagger}
\end{eqnarray}
with  two particles in orbits $a$ and $b$. 
We also introduce the adjoint of $A^\dagger_{JM} (ab)$, the pair annihilation operator,
\begin{eqnarray}
\tilde{A}_{JM} (cd) = - (\tilde{c} \otimes \tilde{d})_{JM}.
\label{tildeA}
\end{eqnarray}
Here we use the standard convention $\tilde{c}_{m_c}= (-1)^{j_c +m_c}\hat{c}_{-m_c}$, where $m_c$ is the $z$-component of angular momentum; this guarantees that if $\hat{a}^\dagger_{jm}$ transforms as a spherical tensor, so does $\tilde{a}_{jm}$ \cite{Edmonds1996Angular}.
An alternate notation is 
\begin{equation}
\hat{A}_{JM}(cd) = \left ( \hat{A}^\dagger_{JM}(cd) \right)^\dagger = (-1)^{J+M} \tilde{A}_{J,-M}(cd).
\end{equation}

With this we can write down a standard form for any one- plus two-body Hamiltonian or Hamiltonian-like operator, which are angular momentum 
scalars.  To simplify we use
\begin{eqnarray}
\hat{H} = \sum_{ab} e_{ab} \hat{n}_{ab} 
+ \frac{1}{4} \sum_{abcd}
\zeta_{ab} \zeta_{cd}
\sum_J V_J(ab,cd) \sum_{M} \hat{A}^\dagger_{JM}(ab) \hat{A}_{JM} (cd) , \label{hamdef}
\end{eqnarray} 
where $\hat{n}_{ab} = \sum_m \hat{a}^\dagger_m \hat{b}_m$ and $\zeta_{ab} = \sqrt{1+\delta_{ab}}$. 
Here $V_J(ab,cd) = \langle a b;J | \hat{V} | cd; J \rangle $ is the matrix element of the purely two-body part of $\hat{H}$ between normalized two-body states with good angular momentum $J$; because $H$ is a 
scalar the value is independent of the $z$-component $M$. 
One can also write this, in slightly different formalism, as
\begin{eqnarray}
\sum_{ab} e_{ab} [j_a] \left ( \hat{a}^\dagger \otimes \tilde{b} \right )_{0,0} 
+ \frac{1}{4} \sum_{abcd} \zeta_{ab} \zeta_{cd}
 \sum_J V_J(ab,cd)\, [J]  \, \left (  \hat{A}^\dagger_{J}(ab) \otimes \tilde{A}_{J} (cd) \right)_{0,0},
 \label{H_scalar}
\end{eqnarray}
where we use the notatation $[x] = \sqrt{2x+1}$, which some authors write as $\hat{x}$ (we use the former to avoid 
getting confused with operators which always are denoted by either $\hat{a}$ or $\tilde{a}$).

Finally we also  introduce one-body transition operators with good angular momentum rank $K$ and $z$-component of angular momentum $M$,
\begin{equation}
\hat{F}_{K,M} = \sum_{ab} F_{ab} [K]^{-1} \left ( \hat{a}^\dagger \otimes \tilde{b} \right)_{K,M} \label{transopdef}.
\end{equation}
Here $F_{ab} = \langle a || \hat{F}_K || b \rangle$ is the reduced one-body matrix element using the Wigner-Eckart theorem and the 
conventions of Edmonds \cite{Edmonds1996Angular}.  For non-charge-changing transitions, Eq.~(\ref{sphtensoradjoint}) implies
\begin{equation}
F_{ab} = (-1)^{j_a -j_b} F_{ba}^*. \label{hermitian_trans}
\end{equation}

With these definitions and conventions, we can now work out general formulas for sum rules.  An important issue will be isospin.  Realistic 
operators, such as M1, 
 connect states with different isospin,
and so rather than working in a formalism with good isospin we treat protons and neutrons 
as being in separate orbits. (Counter to this, we give one example with isoscalar E2 transitions in section \ref{results}.)

\subsection{Non-energy-weighted sum rules}

The non-energy-weighted sum rule operator is given by
\begin{eqnarray}
\hat{O}_{NEWSR} =  \vec{F}^\dagger \cdot \vec{F} =  \sum_M
\left( \hat{F}_{KM}\right)^\dagger \hat{F}_{KM}  
 =  \sum_{M}(-1)^{M} \hat{F}_{K -M} \hat{F}_{KM},
 \end{eqnarray}
 using Eq.~(\ref{sphtensoradjoint}).
Then 
\begin{eqnarray}
\hat{O}_{NEWSR}  &=&
 \sum_{ab} \hat{n}_{ab} \sum_c  \frac{ F_{ca}^* F_{cb} }{2j_a+1} 
-
   \sum_{abcd}
 F_{cb}^* F_{ad}
\sum_J  \left \{
\begin{array}{ccc}
j_a & j_d & K \\
j_c & j_b & J
\end{array}
\right \} 
 \sum_{\mu} \hat{A}^\dagger_{J\mu}(ab) \hat{A}_{J\mu}(cd)  \nonumber\\
 &=&
\label{tensorseparableforce} 
   \sum_{ab} ( \hat{a}^\dagger \otimes \tilde{b} )_{00} \sum_c [j_a]^{-1} F_{ca}^* F_{cb}
 - \sum_{abcd}
 F_{cb}^* F_{ad}
\sum_J  \left \{
\begin{array}{ccc}
j_a & j_d & K \\
j_c & j_b & J
\end{array}
\right \} 
 [J]  \left ( \hat{A}^\dagger_{J}(ab) \otimes  \tilde{A}_{J}(cd) \right )_{00}. \nonumber\\
\end{eqnarray}

By writing out the operator as an angular momentum scalar and to look ``just like'' a Hamiltonian, for purposes of 
use in a shell-model code, we have
\begin{eqnarray}
\hat{O}_{NEWSR} = &\sum_{ab} g_{ab} [j_a ] (a^\dagger \otimes \tilde{b})_{0,0}
 + \frac{1}{4} \sum_{abcdJ} \zeta_{ab} \zeta_{cd}  W_J(ab,cd) \, [J] 
\left( A^\dagger_{J} (ab) \otimes \tilde{A}_{J} (cd) \right)_{0,0},
\nonumber\\
\label{ONEWSR}
\end{eqnarray}
where the single-particle matrix element is
\begin{equation}
g_{ab} = \sum_c \frac{F_{ca}^*F_{cb} }{2j_a +1}. \label{NEWSR1body}
\end{equation} 
We do not assume isospin symmetry, but assume our orbital labels also reference
protons/neutrons. So in (\ref{NEWSR1body}) labels $a$ and $b$ must be the same, proton or neutron.
Now for the two-body matrix elements:
for identical particles in orbits (i.e., $a,b,c,d$ all label  protons or all label neutrons),
we need to enforce antisymmetry, that is,
$W^{pp(nn)}_J(ab,cd) = -(-1)^{j_a + j_b + J}W^{pp(nn)}_J(ba,cd)$, etc:
\begin{eqnarray}
W^{pp(nn)}_J(ab,cd) = - 2\left(1 + \mathscr{P}_{abJ} \right)
\zeta_{ab}^{-1} \zeta_{cd}^{-1} 
\left \{ \begin{array}{ccc}
j_a & j_d & K \\
j_c  & j_b & J 
\end{array} \right \}  F_{cb}^{pp(nn)*} F^{pp(nn)}_{ad},
\end{eqnarray}
where $\mathscr{P}_{abJ}=-(-1)^{j_a+j_b+J}P_{ab}$, and $P_{ab}$ is the exchange operator swapping $a \leftrightarrow b$.
Here the only terms in $\hat{F}$ which contribute are the non-charge-changing pieces, $F^{pp}$ and $F^{nn}$. 

For proton-neutron interactions, where we assume labels $a,c$ are proton and $b,d$ are neutron, i.e., we want 
to compute $W^{pn}_J(a_\pi b_\nu, c_\pi d_\nu)$,  
we need to identify the proton-neutron parts of $\hat{F}$.
So we still have (\ref{NEWSR1body}) and 
\begin{eqnarray}
W^{pn}_J(ab, cd) = - \left(
\left (  F^{pn*}_{cb} F^{pn}_{ad} +(-1)^{j_a+j_b + j_c + j_d}
 F^{np*}_{da} F^{np}_{bc}\right )
  \left \{ \begin{array}{ccc}
j_a & j_d & K \\
j_c & j_b & J 
\end{array} \right\} 
 \nonumber \right . \\
\left .
-(-1)^J \left ( 
(-1)^{j_a +j_b }
 F^{pp*}_{ca} F^{nn}_{bd}+
(-1)^{ j_c + j_d}
 F^{nn*}_{db} F^{pp}_{ac}\right ) 
\left \{ \begin{array}{ccc}
j_a & j_c & K \\
j_d & j_b & J 
\end{array} \right\} 
\right).
\end{eqnarray}
The first two terms are for charge-changing transitions, while the last two are for charge-conserving transitions. Note it is 
possible to create an operator for just one direction, e.g., a non-energy-weighted sum rule for $\beta-$ transitions.  

\subsection{Energy-weighted sum rules}
We define
\begin{eqnarray}
\hat{O}_\mathrm{EWSR} &= & \frac{1}{2}\sum_M (-1)^M \left[ \hat{F}_{K,-M}, [\hat{H}, \hat{F}_{K,M}] \right]  \label{Od.c.} 
\nonumber\\
&= &\sum_{ab} g_{ab} [j_a ] (a^\dagger \otimes \tilde{b})_{0,0}
 + \frac{1}{4} \sum_{abcd} \zeta_{ab} \zeta_{cd} 
\sum_J  W_J(ab,cd) \, [J] 
\left( A^\dagger_{J} (ab) \otimes \tilde{A}_{J} (cd) \right)_{0,0}.
\nonumber\\
\label{Od.c.expansion1}
\end{eqnarray}
In this format the EWSR operator is an angular momentum scalar and, again, looks ``just like'' a Hamiltonian, for purposes of 
use in a shell-model code.

In order to  derive the EWSR as an expectation value of a double-commutator, we \textit{must} use (\ref{sphtensoradjoint}).
  Then, for example, 
for Gamow-Teller we cannot compute the EWSR for $\beta-$ or $\beta+$ alone, but must compute it 
for the sum.  While this is physically less interesting, it is the only possibility for an expectation value of a 
two-body operator.  
 If  we do not use (\ref{sphtensoradjoint}), the EWSR becomes
\begin{equation}
S_1(E_i) =\left  \langle i \left | \hat{F}^\dagger [ \hat{H}, \hat{F} ] \right | i \right \rangle = 
\left  \langle i \left | [\hat{F}^\dagger, \hat{H}] \hat{F}  \right | i \right \rangle
= \frac{1}{2} \left  \langle i \left | \hat{F}^\dagger [ \hat{H}, \hat{F} ]+ [\hat{F}^\dagger, \hat{H}] \hat{F}  \right | i \right \rangle,
\end{equation}
and the resulting operator will have three-body components.

After annihilating commutators and recoupling angular momentums,
the one-body parts of $\hat{O}_{EWSR}$ in Eq.(\ref{Od.c.}) are
\begin{eqnarray}
g_{ab} = \frac{\delta_{j_a j_b} }{2 (2j_a +1)}
\sum_{cd} \left(
-  e_{ac} F_{cd}  F^*_{bd}
+ F_{ac} e_{cd}  F^*_{bd}
+ F^*_{ca} e_{cd} F_{db}
-  F^*_{ca} F_{cd} e_{db}
\right),
\label{g(ab)}
\end{eqnarray}
where $e_{ab}$ are the one-body parts of the Hamiltonian in Eq.(\ref{hamdef}),
and the two-body matrix elements of $\hat{O}_{EWSR}$ are
\begin{eqnarray}
W_J (abcd) = \sum^5_{i=1} W^i (abcd;J),
\end{eqnarray}
with (using Eq.~(\ref{hermitian_trans}) where possible to eliminate or reduce phases)
 \begin{eqnarray}
W^1(abcd;J) &=& - \frac{1}{2} (1 + \mathscr{P}_{cdJ}) \sum_{efJ'} (-1)^{J+J'} (2J'+1) 
\pi^{J^\prime}_{de} \zeta_{ef} \zeta^{-1}_{cd} 
V_J(ab,ef)
\nonumber\\
&& \times F_{ec} F_{fd}
\left\{
\begin{array}{ccc}
J     & K    & J'\\
j_d  & j_e  & j_f
\end{array}
\right\}
\left\{
\begin{array}{ccc}
J     & K    & J'\\
j_e  & j_d & j_c
\end{array}
\right\},
\label{W1}
\\
W^2(abcd;J) &=& - \frac{1}{2} (1 + \mathscr{P}_{cdJ}) \sum_{efJ'} (2J'+1)
\pi^{J^\prime}_{cf} \zeta_{ce} \zeta^{-1}_{cd}
V_J(ab,ce) \nonumber\\
&& \times F_{ef} F^*_{df}
\left\{
\begin{array}{ccc}
J    & K    & J'\\
j_f  & j_c & j_e
\end{array}
\right\}
\left\{
\begin{array}{ccc}
J     &   K  &   J'\\
j_f  &   j_c &  j_d
\end{array}
\right\},
\label{W2}
\\
W^3 (abcd;J) &=& (1+\mathscr{P}_{abJ})(1+\mathscr{P}_{cdJ}) \sum_{efJ'} (2J'+1)
\zeta_{be} \zeta_{df} \zeta^{-1}_{ab} \zeta^{-1}_{cd}
V_{J'}(be,df)
\nonumber\\
&& \times F^*_{ea} F_{fc}
\left\{
\begin{array}{ccc}
J    & K   & J'\\
j_e & j_b & j_a
\end{array}
\right\}
\left\{
\begin{array}{ccc}
J      & K   & J'\\
j_f   & j_d & j_c
\end{array}
\right\},
\label{W3}
\\
W^4(abcd;J) &=& P_{ac} P_{bd} W^{1*}(abcd;J),
\label{W4}
\\
W^5 (abcd;J) &=& P_{ac} P_{bd} W^{2*}(abcd;J),
\label{W5}
\end{eqnarray}
where $\zeta_{ab} = \sqrt{1+\delta_{ab}}$ as former defined, and $\pi^{J^\prime}_{de}$ is defined as
\begin{eqnarray}
\pi^{J^\prime}_{de} = \left\{
\begin{array}{cc}
0, ~~~~~& \text{\rm if} ~~\text{\rm $d=e$ and $J^\prime$ is odd};\\
1, ~~~~~& \text{\rm else}. ~~~~~~~~~~~~~~~~~~~~~~~~
\end{array}
\right.
\label{pi}
\end{eqnarray}
We introduce this symbol because in the derivations of $W^1(abcd;J)$, $J^\prime$ is an intermediate angular momentum, which accounts for the total angular momentum of two fermion annihilators in orbits $d$ and $e$.
As the Pauli principle demands, when $d$ and $e$ are the same orbit, $J^\prime$ must be even in (\ref{W1}). Similarly, in (\ref{W2}) when $c$ and $f$ are the same orbit, $J^\prime$ must be even. 
For detailed explanations please see (\ref{[tildeA,Q]}-\ref{[[tildeA,Q],Q]}) and discussion there. 

\section{Results}
\label{results}

Our formalism applies to configuration-interaction (CI) calculations in a shell-model basis. 
In CI calculations one diagonalizes the many-body Hamiltonian in a finite-dimensioned, orthonormal basis 
of Slater determinants, which are antisymmeterized products of single-particle wavefunctions, typically 
expressed in an occupation representation. The advantage of CI shell model 
calculations is that one can generate excited states easily, and for a modest dimensionality 
one can generate all the eigenstates in the model space.  

We use the {\tt BIGSTICK} CI shell model code \cite{Johnson2013Factorization,BIGSTICK2018} to  calculate the 
many-body matrix elements $H_{\alpha \beta} = \langle \alpha | \hat{H} | \beta \rangle$ and then solve 
$\hat{H} | i \rangle = E_i | i \rangle.$
Greek letters ($\alpha, \beta, \ldots$) denote generic  basis states,
while lowercase Latin letters ($i,j,\ldots$)  label eigenstates. As {\tt BIGSTICK} computes not 
only  energies but also  wavefunctions, we can easily compute sum rules as an expectation value, as in Eq.~(\ref{S_1}). 
We also tested our formalism by fully diagonalizing modest but nontrivial cases, with typical $M$-scheme 
dimensions on the order of a few thousand, where we compute transition density matrices and the subsequent 
transition strengths between all states. This is a straightforward generalization of previous work on the NEWSR \cite{Johnson2015Systematics}.

\begin{figure}[h]
\centering
\includegraphics[width=0.75\textwidth,clip]{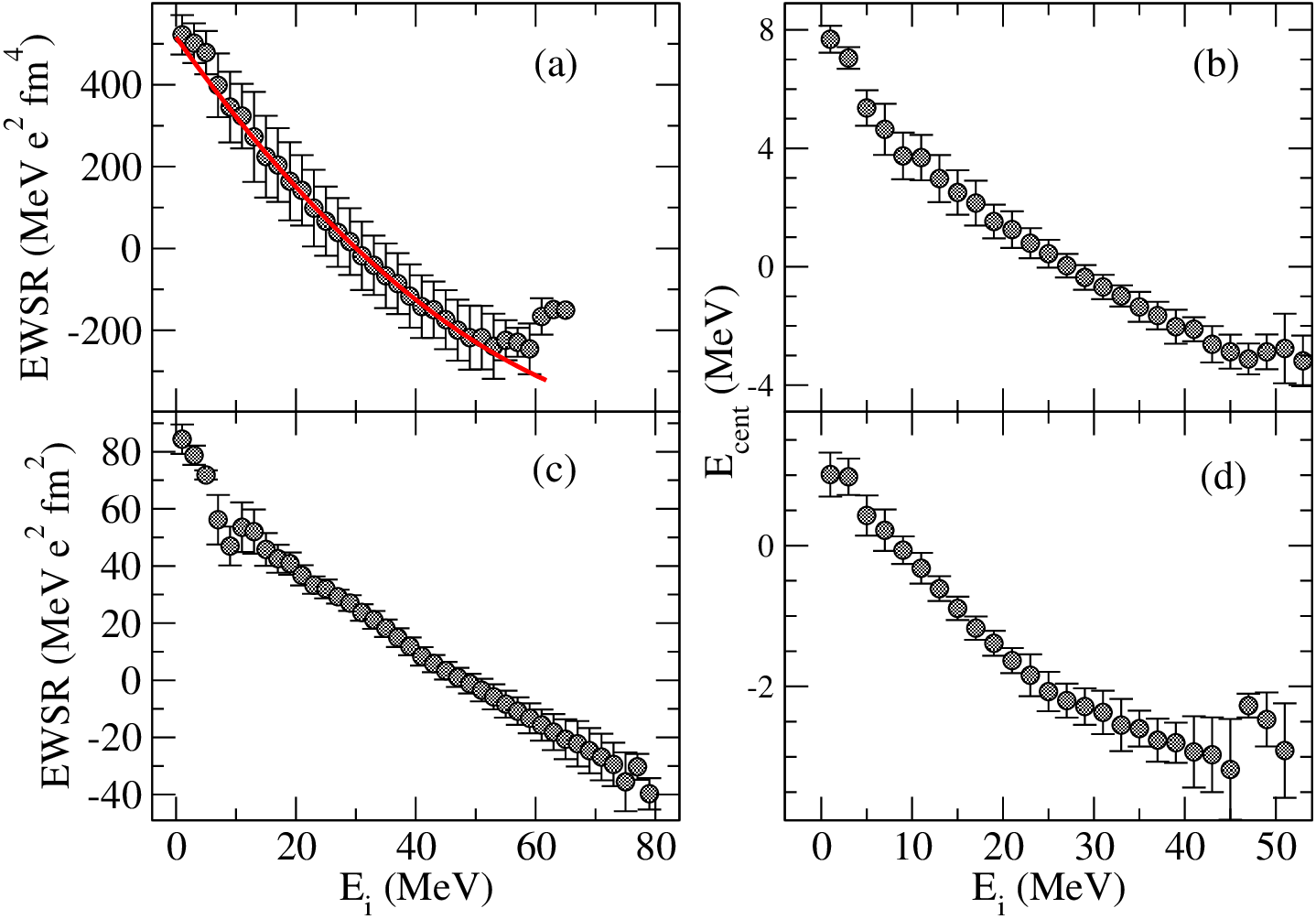}
\caption{ 
 Energy weighted sum rules (EWSR) and transition strength function centroids  as a function of initial energy $E_i$.  
 Results are put into 2 MeV bins with the average and root-mean-square flucutation shown; the fluctuations are not 
 sensitive to the size of the bins.
 (a)  EWSRs for isoscalar E2 
for $^{34}$Cl in the $sd$ shell. The (red) solid line is the secular behavior predicted by spectral distribution theory, as described in Ref. \cite{Johnson2015Systematics}.
(b)  Centroids for M1 transitions 
in $^{21}$Ne in the $sd$ shell. 
 (c)  EWSR for E1 transitions in $^{10}$B in 
$0p$-$1s$-$0d_{5/2}$ space. 
(d) Centroids for Gamow-Teller transitions, sum of $\beta\pm$, for $^{27}$Ne in the $sd$ shell. 
}
\label{all_ewsr}
\end{figure}

To illustrate our formalism we use phenomenological spaces and interactions, for example,
the   $1s_{1/2}$-$0d_{3/2}$-$0d_{5/2}$ or $sd$ shell, 
using a universal $sd$ interaction version 
`B' (USDB)~\cite{PhysRevC.74.034315}.  We show results for selected nuclides, for which we can fully diagonalize the Hamiltonian 
in the model space, as a function of 
initial energy (relative to the ground state)  in Fig.~\ref{all_ewsr}. The centroids are simply evaluated by the ratio of the EWSR to the NEWSR, as 
in Eq.~(\ref{E_centroid}).  
  Because of the finite model space and because we consider the sum rules for \textit{all} states, the centroids and the EWSR must 
go from positive to negative. 
Panel (a) shows  the EWSR for isoscalar E2 transitions in $^{34}$Cl, while panel (b) shows
the centroids for  transitions in $^{21}$Ne with standard $g$-factors \cite{BG77}.
 While we assume harmonic oscillator single-particle wave functions for the basis, taking $\hbar \Omega = 41 A^{-1/3}$MeV, because we compute centroids the oscillator length divides out.  All results were put  into 2 MeV bins, but the 
size of the fluctuations shown by error bars are insensitive to the size of the bins.
 Also shown is the spectral distribution 
theory prediction of the secular behavior:  one  exploits traces of many-body operators to exactly 
arrive at smooth secular behavior shown by the 
red solid line in panel 1(a). 
Not only can one compute the EWSR as an expectation value, 
the secular behavior with excitation energy is quite smooth and by relating the EWSR to the expectation value of an operator, and defining an inner product using many-body traces,
 that behavior can 
be understood from a simple mathematical point of view,  as discussed in more detail in \cite{Johnson2015Systematics} 
 (the reason we choose isoscalar E2 is 
that the publically available code we used to compute the inner product \cite{Launey2014Program} only allows interactions with good isospin).  
Panel (d) shows the centroids for charge-changing Gamow-Teller transitions starting 
from $^{27}$Ne. Because Eq.~(\ref{S_1}) requires the transition operator of rank $K$ to follow (\ref{sphtensoradjoint}), we have to sum both $\beta+$ and $\beta-$ transitions. For $^{27}$Ne the total $\beta-$ strength is 21.239 $g_A^2$, which dominates over $\beta+$ whose 
total strength is 0.239 $g_A^2$, satisfying the Ikeda sum rule.   Again, because we are taking ratios the value of $g_A$ divides out for the centroids.

\begin{figure}
\centering
\includegraphics[scale=0.35,clip]{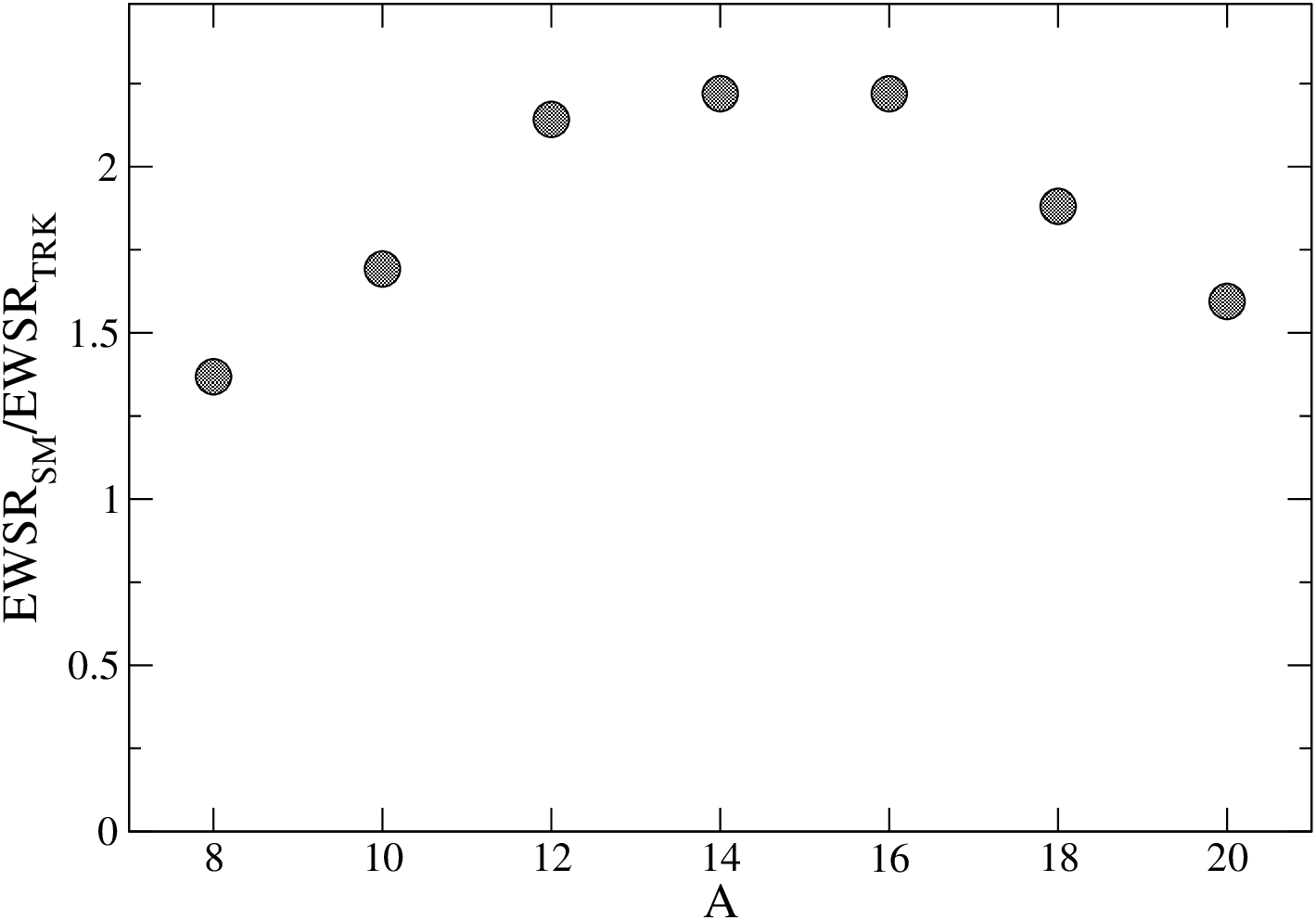}
\caption{ 
Ground state E1 energy-weighted sum rule (EWSR) for $Z=N$ nuclides computed in the $0p$-$1s$-$0d_{5/2}$ shell model space (SM), normalized by 
the Thomas-Reiche-Kuhn (TRK) EWSR. }
\label{e1_ewsr}
\end{figure}

We also considered E1 transitions in  a space with opposite parity orbits, the
$0p_{1/2}$-$0p_{3/2}$-$1s_{1/2}$-$0d_{5/2}$ or $p$-$sd_{5/2}$ space, chosen so we could fully diagonalize for some nontrivial cases. The interactions
 uses the Cohen-Kurath (CK) matrix elements in
the $0p$ shell\cite{Cohen1965Effective}, the older USD interaction \cite{Wildenthal1984Empirical} in
the $0d_{5/2}$-$1s_{1/2}$ space, and the Millener-Kurath (MK)
$p$-$sd$ cross-shell matrix elements\cite{Millener1975The}.  Within the $p$ and
$sd$ spaces  the relative single-particle
energies for the CK and USD interactions, respectively,  are preserved, but 
 $sd$ single-particle energies  shifted relative to the
$p$-shell single particle energies to get the first $3^-$ state 
in $^{16}$O
 at approximately
$6.1$ MeV above the ground state. The rest of the $^{16}$O spectrum, in
particular the first excited $0^+$ state, is not very good, but the
idea is to have a non-trivial model, not exact reproduction of the
spectrum.  Panel (c) of Fig.~\ref{all_ewsr} shows the E1 EWSR  for $^{10}$B, where, as with the other cases, due to the finite model space the sum rule is not 
constant. 
One of the most important and most famous application of sum rules is to electric dipole (E1) transitions, where the Thomas-Reiche-Kuhn (TRK) sum rule \cite{towner1977shell} predicts
$S_1 = (NZ/A) e^2 \hbar^2 / 2m_N$.
Fig.~(\ref{e1_ewsr}) shows the ground state E1 energy-weighted sum rule for $Z=N$ nuclides in this space, normalized by the TRK prediction. 
The enhancement over the TRK sum rule, between 40 and $125\%$, is similar to previous results, \cite{towner1977shell,arima1973effect,weng1973electric,ahrens1975total,PhysRevLett.41.1288,arenhovel1979electromagnetic}. 
While one should not take these results as  
realistic, given the smallness of the model spaces and the crudity of the interaction, it nonetheless illustrates the simplicity of this approach.

\begin{figure}[h]
\centering
\includegraphics[width=0.75\textwidth,clip]{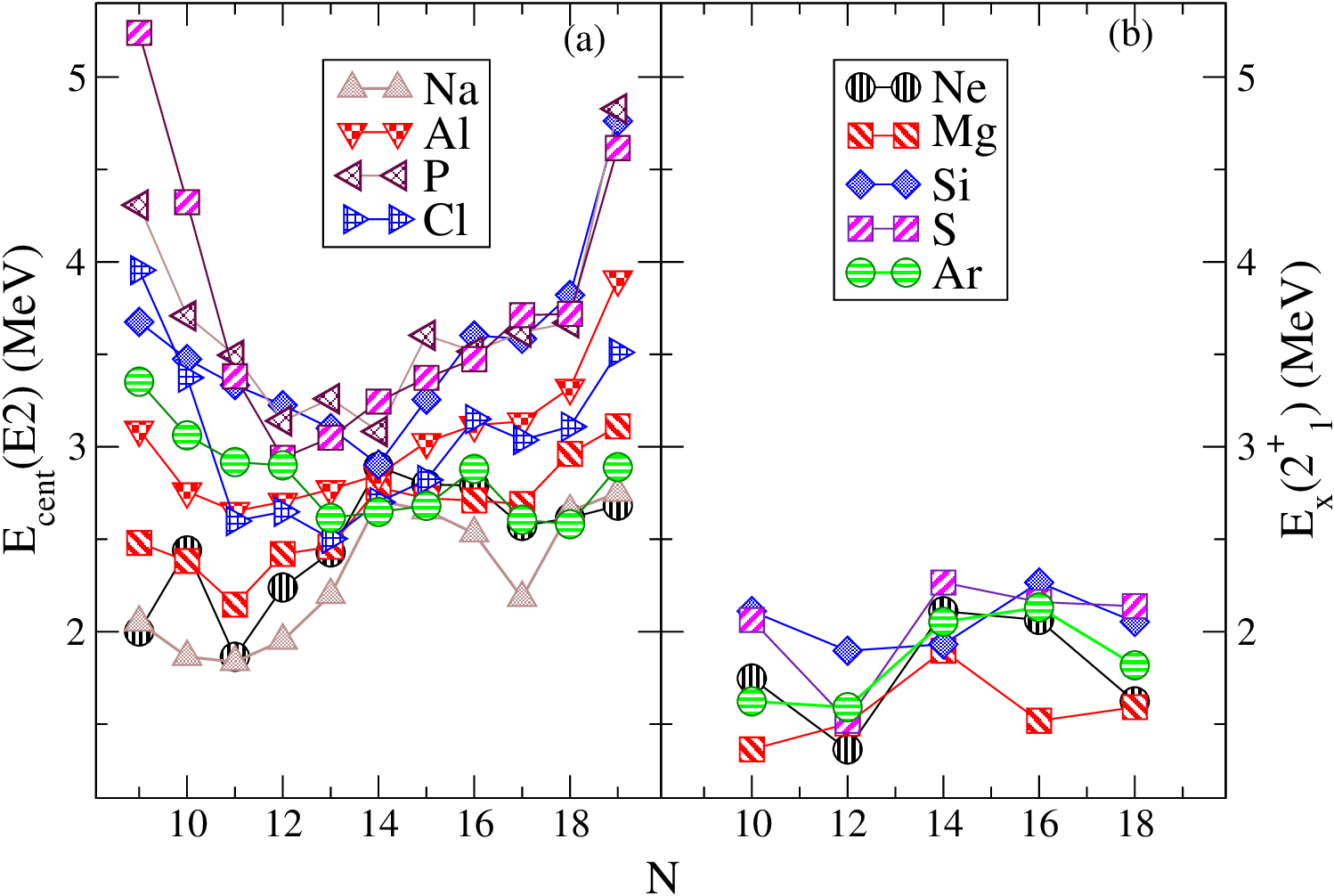}
\caption{ 
In the $sd$-shell using the Brown-Richter USDB interaction. (a) Centroids of E2 transitions from the ground state as a 
function of neutron number $N$. Includes nuclides with both even and odd proton number, with symbols in the boxes 
on both panels. (b) Excitation energies of 
the first $2^+$ state for even-even nuclides only. 
 }
\label{e2centroids}
\end{figure}

By expressing sum rules as operators, one can efficiently search for systematic behaviors. 
For example, we searched for correlations in the $sd$ shell suggested by Eq.~(\ref{M1_EWSR}) but found none. 
Further investigation instead led us to systematics of the $E2$ transitions in the $sd$ shell, shown in Fig.~\ref{e2centroids}. 
Again we used the Brown-Richter USDB interaction, and used effective charges of $1.5e$ and $0.5e$ for protons and neutrons, respectively. 
The USDB interaction is known to be relatively good at producing low-lying energy spectra and transitions of $sd$-shell nuclei, so we use it to calculate $E_{\rm centroid}$; while the $E2$ operator can connect to $2\hbar\Omega$ excitations, such transitions are excluded from this
model space, so here the centroids mostly signal the low-lying transition strengths.
The left panel, (a), gives the energy centroid, the ratio of the EWSR to the NEWSR easily calculated as expectation values, for isotopes of neon through argon, for neutron numbe $N=9$-$19$.  
The data suggest a  convergence at the semi-magic  closure  of the $0d_{5/2}$ shell at $N=14$, which is a maximum for nuclides with 
$Z < 14$ and a minimum for $Z > 14$. 
We have no simple explanation for this behavior, although it seems clearly tied to the semi-magic nature of $N=14$; it is quite different from the 
excitation energy of the first $2^+$ energy in the even-even nuclides, shown in the right panel (b), which, although we do not show it,
closely follow the experimental values. (The closest behavior in the literature we can find are simple behaviors of $2^+_1$ and 
$4^+_1$ excitation energies in heavy nuclei as a function of the number of valence protons and neutrons 
\cite{PhysRevLett.54.1991,PhysRevLett.58.658,PhysRevC.43.R949,brenner1992linearity},
demonstrating the close relationship between collectivity and the proton-neutron interaction. 
However we found that those simple relationships between the number of valence nucleons and the $2^+_1$ and 
$4^+_1$  energies do not hold in the $sd$ shell.) 
 We also note an advantage of sum rules over other regularities such as $2^+$  excitation energies: they can be applied easily to all nuclides, while $E(2^+_1)$ may signal the underlying structure of only even-even nuclei. 
Indeed Fig.~\ref{e2centroids}(a) demonstrates this.   
Clearly much more exploration can be done.

\section{Summary}
\label{sec4}
We presented explicit formulas of operators for non-energy-weighted ($S_0$) and energy-weighted ($S_1$) sums rules of transition strength functions, calculated as expectation values in a shell model occupation-space framework.
These formulas are implemented in the publically available code {\tt PandasCommute} \cite{PandasCommute}, which can  generate  the
sum rule
operator one- and two-body matrix elements from general shell-model interactions and transition operator matrix elements.
We presented examples of  electromagnetic and weak transitions for typical cases in $sd$ and $psd_{5/2}$ shell model spaces; $sd$ shell calculations show that the centroids exhibit an secular dependence on the parent state energy.   Calculation of the E1 energy-weighted sum rule in a 
crude model space nonetheless show an enhancement over the Thomas-Reiche-Kuhn sum rule 
similar to previous results.  We also showed intriguing systematics 
of  E2 centroids in the $sd$ shell.

This methodology can be further extended to no-core shell model spaces, even with isospin non-conserving forces (e.g. Coulomb force).
As one only needs a parent state and the Hamiltonian of the many-body system, $E_{\rm centroid}$ might play the role of a test signal in calculations in sequentially enlarged spaces, thus may be useful to address e.g. quenching, impact of $T=0$/$T=1$ interactions on strength functions and so on.

{\bf Acknowledgement:} 
This material is based upon work supported by the U.S. Department of Energy, Office of Science, Office of Nuclear Physics, 
under Award Number  DE-FG02-96ER40985, and National Natural Science Foundation of China (Grants No. 11225524 and No. 11705100). 
This work is supported in part by the CUSTIPEN (China-U.S. Theory Institute for Physics with Exotic Nuclei)funded by the U.S. Department of Energy, Office of Science under grant number DE-SC0009971, which allowed C. W. Johnson to initiate this collaboration.
Y. Lu is thankful for support to visit San Diego State University for 3 months and the  hospitality extended to him, and with 
 great pleasure thanks Prof. Y. M. Zhao for useful discussions and most kind support.

\appendix
\label{app1}
\section{Derivation of the double commutator}

In this appendix we give some details of the derivation of the matrix elements for the EWSR operator, which 
requires double commutation.  Given the one- and two-body matrix elements of the Hamiltonian, $e_{ab}$ and 
$V_J(ab,cd)$ as defined in (\ref{hamdef}), and the reduced one-body matrix elements $F_{ab}$ of the transition 
operator as in (\ref{transopdef}), we want to find the one-body matrix element $g_{ab}$, and the two-body matrix elements $W_J(ab,cd)$ of the EWSR sum rule operator, as defined in (\ref{Od.c.expansion1}).  We remind the reader that we do not assume isospin symmetry and that the single-particle 
orbit labels, $a,b,c,d$, etc., may refer to distinct proton and neutron orbits. 

Taking the expression of the Hamiltonian in (\ref{H_scalar}) into the double commutator in (\ref{Od.c.}), $\hat{O}_{EWSR}$ splits into two terms,
\begin{eqnarray}
&&\hat{O}_\mathrm{EWSR} 
=  - \frac{1}{2} (-1)^K [K] \sum\limits_{ab} e_{ab} [j_a] \left[  [  \hat{Q}_0(ab), \hat{F}_{K}]_K, \hat{F}_K \right]_0 \nonumber\\
&&	- \frac{1}{8}(-1)^K [K]\sum\limits_{abef} \zeta_{ab} \zeta_{ef} \sum\limits_J V_J(ab,ef) [J] \left[  [ (A^\dagger_J(ab) \otimes \tilde{A}_J(ef))_0, \hat{F}_{K}]_K, \hat{F}_K \right]_0 
\label{Od.c.expansion2} 
\end{eqnarray}
where $\hat{Q}_{KM}(ab)$ is defined as
$\hat{Q}_{KM} (ab) \equiv (\hat{a}^\dagger \otimes \tilde{b})_{KM}$.
We have changed dummy indices in the second term, so that $V_J(ab,ef)$ rather than $V_J(ab,cd)$ appears here, as it does in (\ref{W1}), for convenience of later explanations of how to derive (\ref{W1}).

These terms involve commutators with angular momentum recouplings.
Such commutators are dealt with in a unified manner by authors of Ref. \cite{Chen1993Factorization, Chen1993The} with a generalized Wick theorem. 
We introduce their methodology in brief and return to (\ref{Od.c.expansion2}) with the borrowed tool.
They define a generalized commutator,
\begin{equation}
[\hat{\alpha},\hat{\beta}]=\hat{\alpha}\hat{\beta}-\theta_{\alpha \beta}\hat{\beta}\hat{\alpha},
\label{(a,b)}
\end{equation}
where $\hat{\alpha},\hat{\beta}$ are operators in occupation space, including single-particle fermion creation and annihilation operators,
one-body transition operators, and fermion pair creation and annihilation operators.  If $j_\alpha, j_\beta$ are the angular momenta of 
the operators, then
\begin{equation}
\theta_{\alpha \beta}=
\left\{
\begin{aligned}
-1,&&{j_\alpha, j_\beta~\rm are~half~integers};~~~~\\
1,&&{\rm otherwise}.~~~~~~~~~~~~~~~~~~~
\end{aligned}
\right.
\label{theta_ab}
\end{equation}
With these definitions, it's straight forward to derive
\begin{eqnarray}
[\hat{\alpha} \hat{\beta},\hat{\gamma}]=\hat{\alpha} [ \hat{\beta},\hat{\gamma}]+\theta_{\beta\gamma}[\hat{\alpha},\hat{\gamma}]\hat{\beta}.
\label{ab,c}
\end{eqnarray}
Now we also introduce a generalized commutator with good angular momentum coupling, 
\begin{equation}
[ \hat{\alpha}, \hat{\beta} ]_{jm} 
\equiv ( \hat{\alpha} \otimes \hat{\beta} )_{jm} - (-1)^{j_\alpha + j_\beta - j} \theta_{\alpha \beta} ( \hat{\beta} \otimes \hat{\alpha} )_{jm}.
\end{equation}

and for spherical tensor products
\begin{eqnarray}
&&[(\hat{\alpha} \otimes \hat{\beta})_{j},\hat{\gamma}]_{j^\prime} 
= \sum_{j^{\prime\prime}} U(j_\alpha j_\beta j^\prime j_\gamma ;j j^{\prime \prime})(\hat{\alpha} \otimes [\hat{\beta},\hat{\gamma}]_{j^{\prime \prime}} )_{j^\prime} 
\nonumber\\
&& + \theta_{\beta \gamma}\sum_{j^{\prime \prime}} (-1)^{j_\alpha + j^\prime - j - j^{\prime \prime}} U(j_\alpha j_\beta j_\gamma j^\prime ; j j^{\prime \prime})  
	 ([\hat{\alpha}, \hat{\gamma}]_{j^{\prime \prime}} \otimes \hat{\beta})_{j^\prime},~~~~~~ 
\label{king}
\end{eqnarray}
where
\begin{equation}
U(j_\alpha j_\beta j_\gamma j^\prime ; j j^{\prime \prime}) \equiv (-1)^{j_\alpha+j_\beta+j_\gamma+j^\prime} [j] [j^{\prime \prime} ]
\left\{
\begin{aligned}
j_\alpha ~~ j_\beta ~~ j\\
j^\prime ~~ j_\gamma ~~ j^{\prime \prime}
\end{aligned}
\right\},
\label{U}
\end{equation}
and $[x] \equiv \sqrt{2x+1}$ as defined before.

Now we go back to (\ref{Od.c.expansion2}).
We remind the reader that, according to  (\ref{transopdef}),
$\hat{F}_{K,M} = \sum_{ab} F_{ab} [K]^{-1} \hat{Q}_{K,M}(ab)$, so
{\bf the first term} in (\ref{Od.c.expansion2}) is a linear summation of terms in the form of
$\left[ [ \hat{Q}_0(ab), \hat{Q}_K(cd) ]_K,\hat{Q}_K(ef) \right ]_0$.

With (\ref{king}) we can derive
\begin{eqnarray}
&&\left[ \hat{Q}_J (ab), \hat{Q}_K (cd) \right]_{J^\prime M^\prime } = \left[(\hat{a}^\dagger \otimes \tilde{b})_J, (\hat{c}^\dagger \otimes \tilde{d})_K \right]_{J^\prime M^\prime} 
\nonumber\\
&&= (-1)^{j_a + j_d + J^\prime} \delta_{bc}[J][K] 
\left\{
\begin{aligned}
j_a ~~ j_b ~~ J\\
K ~~ J^\prime~~ j_d
\end{aligned}
\right\}
\hat{Q}_{J^\prime M^\prime}(ad)
\nonumber\\
&& - (-1)^{j_b + j_c + J + K} \delta_{da} [J] [K]
\left\{
\begin{aligned}
j_a ~~ j_b ~~ J\\
J^\prime ~~ K ~~ j_c
\end{aligned}
\right\}
\hat{Q}_{J^\prime M^\prime}(cb),
\label{[Q,Q]}
\end{eqnarray}
and thereafter
\begin{eqnarray}
&&\left [ \left[ \hat{Q}_J(ab), \hat{Q}_K(cd) \right]_{J^\prime}, \hat{Q}_K(ef) \right ]_{JM}
= [J][J^\prime](2K+1) \nonumber\\
&& \left\{ 
 + \phi_{aeK} \delta_{bc} \delta_{fa} 
\left\{
\begin{array}{ccc}
J    &  K  &  J^\prime \\
j_d &  j_a & j_b
\end{array}
\right\}
\left\{
\begin{array}{ccc}
J    &  K  &  J^\prime \\
j_a &  j_d & j_e
\end{array}
\right\}
\hat{Q}_{JM}(ed) \right. \nonumber\\
&&\left.
 - \phi_{dfJJ^\prime} \delta_{bc} \delta_{de}
\left\{
\begin{array}{ccc}
J    &  K  &  J^\prime \\
j_d &  j_a & j_b
\end{array}
\right\}
\left\{
\begin{array}{ccc}
J    &  K  &  J^\prime \\
j_d &  j_a & j_f
\end{array}
\right\}
\hat{Q}_{JM}(af) \right. \nonumber\\
&&\left.
 + \phi_{bfK} \delta_{ad} \delta_{be}
\left\{
\begin{array}{ccc}
J    &  K  &  J^\prime \\
j_c &  j_b & j_a
\end{array}
\right\}
\left\{
\begin{array}{ccc}
J    &  K  &  J^\prime \\
j_b &  j_c & j_f
\end{array}
\right\}
\hat{Q}_{JM}(cf) \right.
\nonumber\\
&&\left.
 - \phi_{ceJJ^\prime} \delta_{ad} \delta_{cf}
\left\{
\begin{array}{ccc}
J    &  K  &  J^\prime \\
j_c &  j_b & j_a
\end{array}
\right\}
\left\{
\begin{array}{ccc}
J    &  K  &  J^\prime \\
j_c &  j_b & j_e
\end{array}
\right\}
\hat{Q}_{JM}(eb) 
\right\}, \nonumber\\
\label{[(Q,Q),Q]}
\end{eqnarray}
where $\phi_{aeK} = (-1)^{j_a + j_e + K}$, other $\phi_{\cdots}$ are similar.
We take (\ref{[(Q,Q),Q]}) into the 1st term in (\ref{Od.c.expansion2}), and end up with the expression for $g_{ab}$ in (\ref{g(ab)}).

{\bf The second term} in (\ref{Od.c.expansion2}) is a linear summation of terms 
$\left[  [ (A^\dagger_J(ab) \otimes \tilde{A}_J(ef))_0, \hat{F}_{K}]_K, \hat{F}_K \right]_0$.
With (\ref{king}) it's straight forward to derive
\begin{eqnarray}
&& \left[ \left( A^\dagger_J(ab) \otimes \tilde{A}_J(ef) \right)_0, \hat{F}_{K} \right]_{K,M}
= \sum\limits_{J^\prime} (-1)^{J+K+J^\prime} [J^\prime] [J]^{-1} [K]^{-1}
\left( A^\dagger_J(ab) \otimes [\tilde{A}_J(ef), \hat{F}_{K}]_{J^\prime} \right)_{K,M}
\nonumber\\
&& + \sum\limits_{J^\prime} [J^\prime] [J]^{-1} [K]^{-1}
\left( [A^\dagger_J(ab), \hat{F}_K]_{J^\prime} \otimes \tilde{A}_J(ef) \right)_{K,M},
\label{AAF}
\end{eqnarray}
and thereafter
\begin{eqnarray}
&&\left[ [ (A^\dagger_J(ab) \otimes \tilde{A}_J(ef))_0, \hat{F}_{K}]_K, \hat{F}_K \right]_0
\nonumber\\
&&= \sum\limits_{J^\prime} [J^\prime] [J]^{-1} [K]^{-1} \left\{
(-1)^{J+K+J^\prime} \left( A^\dagger_J(ab) \otimes \left[ [ \tilde{A}_J(ef), \hat{F}_K]_{J^\prime} , \hat{F}_K \right]_J \right)_0 \right.
\label{A+AFF}
\\
&& \left.+ 2 \left( [A^\dagger_J(ab), \hat{F}_K]_{J^\prime} \otimes [\tilde{A}_J(ef), \hat{F}_K]_{J^\prime} \right)_0
+ (-1)^{J+K+J^\prime} \left( \left[ [A^\dagger_J(ab), \hat{F}_K]_{J^\prime}, \hat{F}_K \right]_J \otimes \tilde{A}_J(ef) \right)_0     \right\}.
\nonumber
\end{eqnarray}
Linear summations of the 1st term in the brace of (\ref{A+AFF}) lead to $W^1(abcd;J)$ and $W^2(abcd;J)$ in (\ref{W1}-\ref{W2}), the 2nd term to $W^3(abcd;J)$ in (\ref{W3}), and the 3rd term to $W^4(abcd;J)$ and $W^5(abcd;J)$ in (\ref{W4}-\ref{W5}).
The symmetry between (\ref{W1}-\ref{W2}) and (\ref{W4}-\ref{W5}) originates from here.

We take the 1st term in the brace of (\ref{A+AFF}) as an example, and explain restrictions caused by Pauli's principle mentioned before.
Use (\ref{king}) again to derive
\begin{eqnarray}
&&\left[ \tilde{A}_{J}(ef), \hat{F}_K \right]_{J^\prime M^\prime} = \sum\limits_{gd} F_{gd} [K]^{-1} \left[\tilde{A}_{J} (ef), \hat{Q}_{K} (gd) \right]_{J^\prime M^\prime} 
\nonumber\\
&&= -  [J] (1+ \mathscr{P}_{efJ})\sum\limits_{d} F_{fd}
\left\{
\begin{aligned}
j_e ~~ j_f ~~ J\\
K ~~ J^\prime ~ j_d
\end{aligned}
\right\}
\tilde{A}_{J^\prime M^\prime}(de).
\label{[tildeA,Q]}
\end{eqnarray}

Based on (\ref{[tildeA,Q]}), we derive 
$\left[ \tilde{A}_{J^\prime}(de), \hat{F}_K \right]_{J M}$ and go further to 
\begin{eqnarray}
&& \left[ [\tilde{A}_{J}(ef), \hat{F}_K]_{J^\prime}, \hat{F}_K \right]_{J,M} 
= \sum\limits_{cdgh}(2K+1)^{-1} F_{gd} F_{hc} \left [ [\tilde{A}_J (ef), \hat{Q}_K (gd) ]_{J^\prime}, \hat{Q}_K (hc) \right ]_{J M}
\nonumber\\
&&= [J] [J^\prime] (1+\mathscr{P}_{efJ}) \sum\limits_{cd} \pi^{J^\prime}_{de} F_{fd} F_{ec}
\left\{
\begin{array}{ccc}
J 	&	K 	& 	J^\prime		\\
j_d & j_e  & j_f
\end{array}
\right\}
\left\{
\begin{array}{ccc}
J 	&  K  &  J^\prime \\
j_e & j_d & j_c
\end{array}
\right\}
\tilde{A}_{JM} (cd)
\label{[[tildeA,Q],Q]}
\\
&& +  (-1)^{J + J^\prime}  [J] [J^\prime] (1+\mathscr{P}_{efJ}) \sum\limits_{cd}\pi^{J^\prime}_{de}  F_{fd} F^*_{cd}
\left\{
\begin{array}{ccc}
J 	&	K 	& 	J^\prime		\\
j_d & j_e  & j_f
\end{array}
\right\}
\left\{
\begin{array}{ccc}
J 	&  K  &  J^\prime \\
j_d & j_e & j_c
\end{array}
\right\}
\tilde{A}_{JM} (ec).
\nonumber
\end{eqnarray}
Note that $\tilde{A}_{J^\prime M^\prime}(de)$ does not show up in (\ref{[[tildeA,Q],Q]}), but as it appeared in (\ref{[tildeA,Q]}) as a necessary stone in the water, therefore the restriction by Pauli's principle on $\tilde{A}_{J^\prime M^\prime}(de)$ is inherited by (\ref{[[tildeA,Q],Q]}), i.e. {\bf when $d$ and $e$ in (\ref{W1}) are the same orbit $J^\prime$ must be even}.
So we introduced $\pi^{J^\prime}_{de}$ as defined in (\ref{pi}) to stand for this restriction.

We take the 1st term of (\ref{[[tildeA,Q],Q]}) into the 1st term in the brace of (\ref{A+AFF}), pick up factors in (\ref{Od.c.expansion2}), and we end up with $W^1(abcd;J)$ in (\ref{W1}); similarly the 2nd term of (\ref{[[tildeA,Q],Q]}) end up with $W^2(abcd;J)$ in (\ref{W2}).
Naturally the restriction $\pi^{J^\prime}_{de}$ is inherited by $W^1(abcd;J)$ and also $W^2(abcd;J)$, but because we exchange indices when deriving $W^2(abcd;J)$, the restriction becomes $\pi^{J^\prime}_{cf}$ in (\ref{W2}).

The same trick is applied to the other two terms in the brace of (\ref{A+AFF}), with (\ref{king}) it's straight forward to derive
\begin{eqnarray}
&&\left[ \hat{A}^\dagger_J(ab), \hat{F}_K \right]_{J^\prime M^\prime}
= \sum\limits_{ef} [K]^{-1} F_{ef} \left[ \hat{A}^\dagger_J(ab), \hat{Q}_K (ef) \right]_{J^\prime M^\prime}
\nonumber\\
&&= (-1)^K [J] (1+ \mathscr{P}_{abJ}) \sum\limits_{e}  F^*_{be}
\left\{
\begin{array}{ccc}
J & K & J^\prime \\
j_e & j_a & j_b
\end{array}
\right\}
\hat{A}^\dagger_{J^\prime M^\prime} (ea),
\label{A+F}
\end{eqnarray}
and thereafter
\begin{eqnarray}
&& \left[ [\hat{A}^\dagger_J(ab), \hat{F}_K], \hat{F}_K \right]_{J M}
\nonumber\\
&& =  [J][J^\prime] (1+ \mathscr{P}_{abJ}) \sum\limits_{eg}\pi^{J^\prime}_{ae} F^*_{be} F^*_{ag} 
\left\{
\begin{array}{ccc}
J & K & J^\prime \\
j_e & j_a & j_b
\end{array}
\right\}
\left\{
\begin{array}{ccc}
J & K & J^\prime \\
j_a & j_e & j_g
\end{array}
\right\}
\hat{A}^\dagger_{JM}(ge)
\nonumber\\
&& + (-1)^{J+J^\prime} [J][J^\prime]  (1+ \mathscr{P}_{abJ}) \sum\limits_{eg}\pi^{J^\prime}_{ae} F^*_{be} F_{ge} 
\left\{
\begin{array}{ccc}
J & K & J^\prime \\
j_e & j_a & j_b
\end{array}
\right\}
\left\{
\begin{array}{ccc}
J & K & J^\prime \\
j_e & j_a & j_g
\end{array}
\right\}
\hat{A}^\dagger_{JM}(ag).
\label{A+FF}
\end{eqnarray}
With (\ref{[tildeA,Q]}, \ref{A+F}) one can derive the second term in the brace of (\ref{A+AFF}), and end up with $W^3(abcd;J)$ in (\ref{W3});
with (\ref{A+FF}) one can derive the third term in the brace of (\ref{A+AFF}), and get $W^4(abcd;J)$ and $W^5(abcd;J)$ in (\ref{W4}-\ref{W5}) after picking up factors in (\ref{Od.c.expansion2}).

\bibliographystyle{apsrev4-1}
\bibliography{sumrule}

\end{document}